\documentclass[preprint]{ptephy_v1}
\usepackage{amssymb}
\usepackage{graphicx}
\numberwithin{equation}{section}

\newcommand{\dl}{\delta}
\newcommand{\cb}{\bar{c}}
\newcommand{\lb}[1]{\label{#1}}
\newcommand{\cl}{\mathcal{L}}
\newcommand{\bb}{\bar{B}}

\newcommand{\vp}{\varphi}
\newcommand{\ptl}{\partial}

\def\vec#1{\mbox{\boldmath $#1$}}

\newcommand{\mcb}{\mathcal{B}}

\begin{document}

\title{Confining potential under the gauge field condensation 
in the SU(2) Yang-Mills theory
}


\author{Hirohumi Sawayanagi\thanks{National Institute of Technology, Kushiro College, Kushiro, 084-0916, Japan\\\quad E-mail:sawa@kushiro-ct.ac.jp}}


\begin{abstract}
     $Q\bar{Q}$ potential is studied in the  SU(2) gauge theory.  
Based on the nonlinear gauge of the Curci-Ferrari type, the possibility of 
a gluon condensation $\langle A_{\mu}^+A_{\mu}^-\rangle$ in low-energy region has been considered 
at the one-loop level.  
Instead of the magnetic monopole condensation, 
this condensation makes classical gluons massive, and can yield a linear potential.  
We show this potential consists of the Coulomb plus linear part and an 
additional part.  Comparing with the Cornell potential, we study this confining potential in detail, 
and find that the potential has two implicit scales $r_c$ and $\tilde{R}_0$.  The meanings of 
these scales are clarified.  We also show that the Cornell potential that fits well to 
this confining potential is obtained by taking these scales into account. 
\end{abstract}

\subjectindex{B0,B3,B6}

\maketitle

\section{Introduction}

     In the study of quarkonia, QCD potential is often used.  Although there are some phenomenological potentials (see, e.g., \cite{bali}), 
the Cornell potential $V_{CL}(r)$ \cite{cor, cor2} is simple but workable.  This potential has the Coulomb plus linear form as 
\begin{equation}
  V_{CL}(r) = -\frac{K}{r} + \sigma r ,   \lb{101}
\end{equation}
where $K$ and $\sigma$, that is called the string tension, are constants.  The Coulomb part is expected from the 
perturbative one-gluon exchange, and the linear part represents the confinement.

     Is it possible to derive $V_{CL}(r)$ from QCD?  Using the dual Ginzburg-Landau model (see, e.g., \cite{rip}), 
the following Yukawa plus linear potential was obtained \cite{suz, mts, sst, sst2}:
\begin{equation}
  V_{YL}(r) = -\frac{Q^2}{4\pi}\frac{e^{-mr}}{r} + \left(\frac{Q^2m^2}{8\pi}\ln\frac{m^2+m_{\chi}^2}{m^2} \right) r,    \lb{102}
\end{equation}
where $Q$ is the static quark charge and $m_{\chi}$ is the momentum cut-off.  In this model, the mass $m$ 
is related to the vacuum expectation value (VEV) of the monopole field.  
In Ref.~\cite{hs}, based on the SU(2) gauge theory in the non-linear gauge of the Curci-Ferrari type, we also derived the potential $V_{YL}(r)$.  
In this case, the mass $m$ comes from the gauge field condensation $\langle A_{\mu}^+A_{\mu}^-\rangle$.

     In this paper, in the framework of Refs.~\cite{hs, hs1}, we restudy the confining potential.  
In the next section, we briefly review Refs.~\cite{hs,hs1}, and present the potential between the static charges $Q$ and $\bar{Q}$.  
In Sect.~3, the equation to determine an ultraviolet cut-off $\Lambda_c$ is derived.  
In Sect.~4, using this cut-off, we show that the $Q\bar{Q}$ potential becomes the confining potential $V_c(r)=V_{CL}(r)+V_3(r)$, 
where $V_3(r)$ is the additional potential.  
The potential $V_c(r)$ has several parameters.  
Comparing $V_c(r)$ with $V_{CL}(r)$, and choosing the appropriate values of $K$ and $\sigma$, the parameters in $V_c(r)$ are determined in Sect.~5.  
In this process, we find a scale $r_c \approx 0.2\ \mathrm{fm}$.  
In the intermediate region, the scale $R_0\approx 0.5\ \mathrm{fm}$ has been proposed \cite{so}.  The meanings of the scales $r_c$ and $R_0$ for $V_c(r)$ 
are clarified in Sect.~6.  We also propose a scale $\tilde{R}_0$ that is related to $R_0$.  
Based on this analysis, we obtain $V_{CL}(r)$ that fits well to $V_c(r)$.  
Section~7 is devoted to a summary and comments.  
In Appendix~A, the propagator for the off-diagonal gluons is presented.  The equations in Sect.~5, that determine the values 
of the parameters in $V_c(r)$, are solved in Appendix~B.

\section{Condensate $\langle A_{\mu}^+A_{\mu}^-\rangle$ and $Q\bar{Q}$ potential}

      In this section, we review Refs.~\cite{hs,hs1} briefly.  

\subsection{Ghost condensation}

     We consider the SU(2) gauge theory in Euclidean space.  The Lagrangian in 
the nonlinear gauge of the Curci-Ferrari type \cite{cf} is 
\begin{align}
 \cl &= \cl_{inv}+ \cl_{NL},\quad \cl_{inv}=\frac{1}{4}F_{\mu\nu}^2, \notag \\
  \cl_{NL} &=  B\cdot \partial_{\mu}A_{\mu}+i\cb\cdot\partial_{\mu}D_{\mu}c-
 \frac{\alpha_1}{2}B^2-\frac{\alpha_2}{2}\bb^2 -B\cdot w,   \lb{201}
\end{align}
where $B$ is the Nakanishi-Lautrup field, $c$ ($\bar{c}$) is the ghost (antighost), 
$\bb = -B+ ig\cb \times c$, $\alpha_1$ and $\alpha_2$ are gauge parameters, and 
$w$ is a constant to keep the BRS symmetry.  
Introducing the auxiliary field $\varphi$, which represents $\alpha_2 \bb$, $\cl_{NL}$ 
is rewritten as \cite{hs2} 
\begin{equation}
 \cl_{\vp}=-\frac{\alpha_1}{2}B^2 
 +B\cdot (\ptl_{\mu}A_{\mu}+\vp -w)+i\cb \cdot(\ptl_{\mu}D_{\mu}+g\vp \times )c 
 +\frac{\vp^2}{2\alpha_2}.  \lb{202}
\end{equation}
In Ref.~\cite{hs3}, by integrating out $\bar{c}$ and $c$ with momentum $\mu\leq p\leq \Lambda$, 
we studied the one-loop effective potential for $\vp$, and showed that 
$g \vp^A$ acquires the VEV $g\vp_0\delta^{A3}$ under an energy scale $\mu_0$.  
The scale $\mu_0$ and the VEV $v=g\vp_0$ are 
\begin{equation}
 \mu_0=\Lambda e^{-4\pi^2/(\alpha_2g^2)},\ 
 v=\left\{\frac{\mu_0^4-\mu^4}{1-e^{-16\pi^2/(\alpha_2g^2)}}\right\}^{1/2}.     \lb{203}
\end{equation}
At the one-loop order, it is shown that $\alpha_2= \beta_0/2$ is the ultraviolet fixed point, where 
$\beta_0=22/3$ is the first coefficient of the renormalization group function $\beta$ for SU(2).  
So, when $\Lambda \gg \mu_0 \gg \mu$, the relation 
\begin{equation}
      \mu_0=\Lambda_{\mathrm{QCD}}, \quad v\simeq \Lambda_{\mathrm{QCD}}^2   \lb{204} 
\end{equation}
holds \cite{hs3}, where $\Lambda_{\mathrm{QCD}}$ is the QCD scale parameter.

\subsection{Condensate $\langle A_{\mu}^+A_{\mu}^-\rangle$}

     When $v\neq 0$, the ghost Lagrangian becomes 
\begin{equation}
     i\sum_{a=1}^2 \left(\cb^a\Box c^a-\sum_{b=1}^2v\epsilon_{ab}\cb^ac^b\right) + i\cb^3 \Box c^3.  \lb{205}
\end{equation}
Because of the $v$ term in Eq.(\ref{205}), the two-point function 
\[ \langle A_{\mu}^+(y)A_{\nu}^-(x)\rangle =G_{\mu\nu}(y,x) \] 
shows the tachyonic behavior.  
In fact, the ghost loop depicted in Fig.~1(b) yields the tachyonic mass $(-g^2v)/(64\pi)$ for $A_{\mu}^a\ (a=1,2)$ in 
the low momentum limit $p\to 0$.  In the same way, $A_{\mu}^3$
has the tachyonic mass $-g^2v/(32\pi)$ in this limit.

\begin{figure}
\begin{center}
\includegraphics{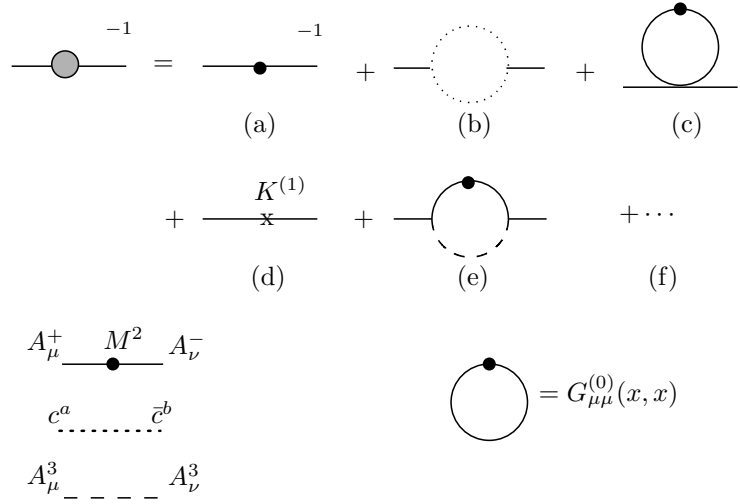}
\caption{The diagrams that contribute to the inverse propagator for $A_{\mu}^{\pm}$.  Fig.~1(b) yields the tachyonic mass in the 
limit $p \to 0$, and Fig.~1(c) brings about the VEV $G_{\mu\mu}^{(0)}(x,x)$.}
\label{fig1}
\end{center}
\end{figure}

     To remove the tachyonic mass terms 
\begin{equation}
   \frac{1}{2}\left(\frac{-g^2v}{64\pi}\right)\left\{\sum_{a=1}^2(A_{\mu}^a)^2+2(A_{\mu}^3)^2\right\}, \lb{206}
\end{equation}
we introduce the source term 
\[ KA_{\mu}^+A_{\mu}^-, \quad K=K^{(0)}+K^{(1)}+\cdots, \quad K^{(n)}=O(\hslash^n).  \]
Although the source $K$ may depend on the momentum scale, for simplicity, 
we treat it as constant, and write $K^{(0)}=M^2$.  
To consider the inverse propagator for $A_{\mu}^{\pm}$ depicted in 
Fig.~1 in the limit $p \to 0$, we write the free part of the propagator $G_{\mu\nu}(y,x)$ as $G_{\mu\nu}^{(0)}(y,x)$.   
Then the diagram in Fig.~1(c) gives the VEV $g^2G_{\mu\mu}^{(0)}(x,x)$.  
If $K^{(1)}$ subtracts divergent terms of $O(\hslash)$ in this limit, 
the condition 
\begin{equation}
     G_{\mu\mu}^{(0)}(x,x) = \frac{v}{64\pi},    \lb{207}
\end{equation}
removes the tachyonic mass for $A_{\mu}^{\pm}$.  Because of the interaction $g^2(A_{\mu}^+A_{\mu}^-)(A_{\nu}^3)^2$ 
in $F_{\mu\nu}^2/4$, this VEV also removes the tachyonic mass for $A_{\mu}^3$.

\subsection{Inclusion of a classical solution}

     To introduce a classical solution, the gauge field $A_{\mu}^A$ is divided into 
the classical part $b_{\mu}^A=b_{\mu}\delta^{A3}$ and the quantum part $a_{\mu}^A$, i.e., 
$A_{\mu}^{\pm}=a_{\mu}^{\pm},\ A_{\mu}^3=b_{\mu}+a_{\mu}^3$.  As the tachyonic masses come from 
the ghost loops with the VEV $v$, it is expected that $b_{\mu}$ acquires no tachyonic mass.  
To see it, we divide the gauge transformation $\delta A_{\mu} = D_{\mu}(A) \varepsilon$ 
as 
\[   \delta a_{\mu} = D_{\mu}(a) \varepsilon, \quad  \delta b_{\mu} = gb_{\mu}\times \varepsilon,  \]
where $D_{\mu}(A)=(\ptl_{\mu}+gA_{\mu}\times)$.  Then, using the gauge-fixing function 
$G=\ptl_{\mu}a_{\mu}+ \vp-w$, the ghost Lagrangian 
\[  B\cdot [\ptl_{\mu}a_{\mu}+\vp-w ]+i\cb \cdot[\ptl_{\mu}D_{\mu}(a)+g\vp \times ]c  \]
is obtained.  If $A_{\mu}$ is replaced by $a_{\mu}$, Eq.(\ref{202}) becomes this Lagrangian.  
So, the tachyonic mass terms for $a_{\mu}^A$ are 
\[
   \frac{1}{2}\left(\frac{-g^2v}{64\pi}\right)\left\{\sum_{a=1}^2(a_{\mu}^a)^2+2(a_{\mu}^3)^2\right\}, 
\]
and $b_{\mu}$ acquires no tachyonic mass.  
\footnote{In Ref.~\cite{hs1}, using the background covariant gauge-fixing, we showed, although 
the ghost Lagrangian contains $b_{\mu}$, it does not acquire the 
tachyonic mass.  } 

     Next, we consider the effect of the VEV $G_{\mu\mu}^{(0)}(x,x)$.  As in the previous subsection, 
these tachyonic mass terms for $a_{\mu}^{A}$ are removed by the VEV in Eq.(\ref{207}).  In addition, 
the interaction 
$g^2(A_{\mu}^+A_{\mu}^-)(a_{\nu}^3+b_{\nu})^2$ in $F_{\mu\nu}^2/4$ generates the 
mass term 
\begin{equation}
     \frac{m^2}{2}(b_{\nu})^2,\quad m^2=2g^2G_{\mu\mu}^{(0)}(x,x).  \lb{208}
\end{equation}
Thus, after integrating out $c$ and $\bar{c}$, an effective  low energy Lagrangian becomes 
\begin{align}
     \cl =& \frac{1}{4}(\ptl \wedge b)_{\mu\nu}(\ptl \wedge b)_{\mu\nu} + \frac{m^2}{2}b_{\mu}b_{\mu}
     +\frac{1}{4}(\ptl \wedge a^3)_{\mu\nu}(\ptl \wedge a^3)_{\mu\nu}  \nonumber \\ 
     & +\frac{1}{2}(\ptl \wedge A^+)_{\mu\nu}(\ptl \wedge A^-)_{\mu\nu} +M^2A^+_{\mu}A^-_{\mu}+ \cdots,  \lb{209}
\end{align}
where $(\ptl \wedge A)_{\mu\nu}=\ptl_{\mu}A_{\nu}-\ptl_{\nu}A_{\mu}$.  
Namely, although the quantum part $a_{\mu}^3$ is massless, the classical part $b_{\mu}$ has the mass $m$.  
The off-diagonal components 
$A_{\mu}^{\pm}$ have the mass $M$ determined by the equation 
\begin{equation}
     \frac{v}{64\pi}=  G_{\mu\mu}^{(0)}(x,x).  \lb{210}
\end{equation}

\subsection{$Q\bar{Q}$ potential}

     Now we consider the confining potential.  As the classical field $b_{\mu}$, 
we choose the dual electric potential $\mcb_{\mu}$, that describes the electric monopole solution \cite{hs}.  
The color electric current $j_{\mu}$ is incorporated by the replacement 
\[ (\ptl \wedge \mcb)_{\mu\nu}  \to 
 (\ptl \wedge \mcb)_{\mu\nu} + \epsilon_{\mu\nu\alpha\beta}(n\cdot\ptl)^{-1}n_{\alpha}j_{\beta},  \]
where the space-like vector $n_{\mu}$ \cite{zwa} is chosen as $n_{\mu}=(0,\vec{n})$ with $\vec{n}\cdot\vec{n}=1$, and 
$n\cdot \ptl=n_{\mu}\ptl_{\mu}$.  
We note this is the Zwanziger's dual field strength $F^d=(\ptl \wedge B)+(n \cdot \ptl)^{-1}(n \wedge j_e)^d$ in Ref.~\cite{zwa}.  
Thus the classical part of $\cl$ in Eq.(\ref{209}) becomes 
\begin{equation}
   \frac{1}{4}\left[(\ptl \wedge \mcb)_{\mu\nu} + \epsilon_{\mu\nu\alpha\beta}(n\cdot\ptl)^{-1}n_{\alpha}j_{\beta}\right]^2
   + \frac{m^2}{2}\mcb_{\mu}\mcb_{\mu}.  \lb{211}
\end{equation}
The equation of motion for $\mcb_{\mu}$ is 
\[   (D_m^{-1})_{\mu\nu}\mcb_{\nu}= \epsilon_{\mu\rho\alpha\beta}(n\cdot\ptl)^{-1}n_{\rho}\ptl_{\alpha}j_{\beta},\quad 
(D_m^{-1})_{\mu\nu}=(-\square +m^2)\delta_{\mu\nu} + \ptl_{\mu}\ptl_{\nu},  \] 
and $\mcb_{\mu}$ is solved as 
\begin{equation}
         \mcb_{\mu}=(D_m)_{\mu\nu}\epsilon_{\nu\rho\alpha\beta}(n\cdot\ptl)^{-1}n_{\rho}\ptl_{\alpha}j_{\beta}, \quad 
(D_m)_{\mu\nu}=\frac{\delta_{\mu\nu}-\ptl_{\mu}\ptl_{\nu}/\square}{-\square+m^2}+\frac{\ptl_{\mu}\ptl_{\nu}}{m^2\square}.  \lb{212}
\end{equation}
If we use Eq.(\ref{212}), Eq.(\ref{211}) becomes 
\begin{equation}
 \cl_{jj}=
 \frac{1}{2}j_{\mu}\frac{1}{-\square + m^2}j_{\mu} 
   -\frac{1}{2}j_{\mu}\frac{m^2}{-\square + m^2}\frac{n\cdot n}{(n\cdot\ptl)^2}
\left(\delta_{\mu\nu}-\frac{n_{\mu}n_{\nu}}{n\cdot n}\right)j_{\nu}.   \lb{213}
\end{equation}

     To derive the static potential between the color electric charges $Q$ and $-Q$, 
the static current 
\begin{equation}
     j_{\mu}(x)=Q\delta_{\mu 0}\{\delta(\vec{x}-\vec{a})-\delta(\vec{x}-\vec{b})\}  \lb{214}
\end{equation}
is substituted into $\cl_{jj}$.  Then it leads to the potential 
\begin{align}
     V(r)=& V_{Y}(r)+V_{L}(r),\quad V_{Y}=Q^2\int \frac{d^3q}{(2\pi)^3}\frac{1- \cos \vec{q}\cdot \vec{r}}{q^2+m^2}, \nonumber \\ 
     & V_{L}=Q^2\int \frac{d^3q}{(2\pi)^3}(1- \cos \vec{q}\cdot \vec{r})\frac{m^2}{(q^2+m^2)q_n^2},  \lb{215}
\end{align}
where $\vec{r}=\vec{a}-\vec{b}$, $q=|\vec{q}|$ and $q_n=\vec{q}\cdot\vec{n}$.  Although the term $V_{Y}$ ($V_{L}$) becomes the Yukawa (linear) potential 
in Eq.(\ref{102}) usually, we restudy Eq.(\ref{215}) in Sect.~4.

\section{Cut-off $\Lambda_c$ and the mass $M$}

     In this section, we study Eq.(\ref{210}).  The free propagator $G_{\mu\nu}^{(0)}(p)$ is calculated in Appendix A as 
\begin{equation}
     G_{\mu\nu}^{(0)}(p)=\frac{1}{p^2+M^2}P^T_{\mu\nu} + \frac{1}{M^2}P^L_{\mu\nu},\quad
     P^T_{\mu\nu}=\delta_{\mu\nu}-\frac{p_{\mu}p_{\nu}}{p^2},\quad P^L_{\mu\nu}=\frac{p_{\mu}p_{\nu}}{p^2}.  \lb{301}
\end{equation}
Assuming that $M$ exists below a cut-off $\Lambda_c$, we obtain 
\begin{equation}
    G_{\mu\mu}^{(0)}(x,x) =\int_0^{\Lambda_c}\frac{d^4p}{(2\pi)^4} G_{\mu\mu}^{(0)}(p)=
 \frac{1}{(4\pi)^2}\left\{ \frac{\Lambda_c^4}{2M^2}-3M^2\ln\left(1+\frac{\Lambda_c^2}{M^2}\right) \right\},  \lb{302}
\end{equation}
where the $M$-independent term $\Lambda_c^2$ is subtracted.

     The VEV $v$ depends on the momentum scale $\mu$.  From Eqs.(\ref{203}) and (\ref{204}), we find 
$v$ disappears above the scale $\mu_0=\Lambda_{\mathrm{QCD}}$.  
When $0\leq \mu \leq \Lambda_{\mathrm{QCD}}$, $v$ behaves as 
\[ v \to 0 \quad (\mu\to \Lambda_{\mathrm{QCD}}), \quad  v\to \Lambda_{\mathrm{QCD}}^2 \quad (\mu\to 0).  \]
Namely the maximal value of $v$ is $\Lambda_{\mathrm{QCD}}^2$, and the left hand side of Eq.(\ref{210}) satisfies 
\[ 0\leq \frac{v}{64\pi} \leq \frac{\Lambda_{\mathrm{QCD}}^2}{64\pi}.  \]
On the other hand,  the VEV $G_{\mu\mu}^{(0)}(x,x) $ only depends on the constants $M$ and $\Lambda_c$.  
Since the tachyonic mass should be removed in the overall momentum region completely, using the maximal value of $v$, 
we interpret Eq.(\ref{210}) as 
\footnote{If we consider the scale dependent $M$,  Eq.(\ref{302}) should be replaced by the integral 
\[ \int_0^{\Lambda_c} \frac{d^4p}{(2\pi)^4}\left\{\frac{P_{\mu\nu}^T}{p^2+M(p)^2}+\frac{P_{\mu\nu}^L}{M(p)^2}\right\}. \] 
As $M(p)$ is unknown, we cannot calculate it.  But it is also $\mu$-independent. }

\begin{equation}
   \frac{\Lambda_{\mathrm{QCD}}^2}{64\pi}\simeq  G_{\mu\mu}^{(0)}(x,x).  \lb{303}
\end{equation}
From Eqs.(\ref{302}) and (\ref{303}), we obtain 
\begin{equation}
     \frac{\pi}{2}\frac{\Lambda_{\mathrm{QCD}}^2}{M^2}\simeq \frac{\Lambda_c^4}{M^4}-6\ln\left(1+\frac{\Lambda_c^2}{M^2}\right).  \lb{304}
\end{equation}

     We make a comment.  
Eq.(\ref{304}) determines the relation of $M$ and $\Lambda_c$ so as to give the maximal value of $v$.  
Using Eq.(\ref{303}), we find $m^2=2g^2 G_{\mu\mu}^{(0)}(x,x)\simeq g^2\Lambda_{\mathrm{QCD}}^2/(32\pi)$.  
Although the VEV $v$ depends on the scale $\mu$, and disappears above $\Lambda_{\mathrm{QCD}}$, 
$G_{\mu\mu}^{(0)}(x,x)$ does not.  So the mass $m$ can survive above the scale 
$\Lambda_{\mathrm{QCD}}$.

\section{Confining potential}

     Usually, the potential $V$ in Eq.(\ref{215}) is calculated as follows.  Let us divide the momentum $\vec{q}$ in $V_L$ into 
$q_n=\vec{q}\cdot\vec{n}$ and $\vec{q}_T$ that satisfies $\vec{q}_T\cdot \vec{n}=0$.  Then the integral of $q_n$ has 
the infrared divergence, and the integral of $q_T=|\vec{q}_T|$ has the ultraviolet divergence.  
The former divergence is removed by the choice $\vec{r} \parallel \vec{n}$ \cite{sst, hs}, and the latter divergence is 
avoided by the cut-off $m_{\chi}$ as \cite{suz, mts, sst, sst2} 
\begin{equation}
     V_{L}=\frac{Q^2m^2}{(2\pi)^2}\int_{-\infty}^{\infty} dq_n\int_0^{m_{\chi}}dq_T q_T\frac{1- \cos q_n r}{(q_n^2+q_T^2+m^2)q_n^2}. \lb{401}
\end{equation}
Eq.(\ref{401}) becomes the linear term in Eq.(\ref{102}).  The term $V_Y$ has the integral of $q=|\vec{q}|$ 
over the region of $0\leq q<\infty$, and the Yukawa potential in Eq.(\ref{102}) is obtained.

     To introduce a cut-off in a different way, we write the potential as 
\begin{equation}
     V(r)=\int dq W(\vec{q},m,r).  \lb{402}
\end{equation}
As we stated in Sect.~3, the mass $m$ can exist above $\Lambda_{\mathrm{QCD}}$.  
From Fig.~1(c), after $v$ disappears, it is expected that 
$m^2$ contributes to the correction for $M^2$.  
Since $A_{\mu}^{\pm}$ are considered to be massless above $\Lambda_c$, 
we assume that the cut-off for $m$ 
is $\Lambda_c$, and define Eq.(\ref{402}) as 
\begin{equation}
     V(r)= \int_0^{\Lambda_c} dq W(\vec{q},m,r) + \int_{\Lambda_c}^{\infty} dq W(\vec{q},0,r).  \lb{403}
\end{equation}
Eq.(\ref{403}) is rewritten as 
\begin{equation}
     V(r)= \int_{0}^{\infty} dq W(\vec{q},0,r)
    + \int_0^{\Lambda_c} dq \left\{W(\vec{q},m,r) -W(\vec{q},0,r)\right\}.  \lb{404}
\end{equation}
The first term becomes 
\begin{equation}
\int_{0}^{\infty} dq W(\vec{q},0,r)=V_1(r)=
 Q^2\int_{D_{\infty}} \frac{d^3q}{(2\pi)^3}\frac{1- \cos \vec{q}\cdot \vec{r}}{q^2}, \quad
D_{\infty}=\{q\ |\ 0\leq q<\infty \}, \lb{405}
\end{equation}
and the second term leads to 
\begin{align}
  &\int_0^{\Lambda_c} dq \left\{W(\vec{q},m,r) -W(\vec{q},0,r)\right\}= V_2(r) + V_3(r), \nonumber \\
 &V_2(r)= Q^2\int_{D_{\Lambda_c}} \frac{d^3q}{(2\pi)^3}(1- \cos \vec{q}\cdot \vec{r})\frac{m^2}{(q^2+m^2)q_n^2}, \nonumber \\
 &V_3(r)= -Q^2\int_{D_{\Lambda_c}} \frac{d^3q}{(2\pi)^3}(1- \cos \vec{q}\cdot \vec{r})\frac{m^2}{(q^2+m^2)q^2}, \quad 
 D_{\Lambda_c}=\{q\ |\ 0\leq q<\Lambda_c \},  \lb{406}
\end{align}
where $V_2(r)$ ($V_3(r)$) comes from $V_L$ ($V_Y(m^2)-V_Y(m^2=0)$).  Eq.(\ref{405}) gives the usual Coulomb potential 
\begin{equation}
     V_1(r)=-\frac{K_c}{r},\quad K_c=\frac{Q^2}{4\pi}.  \lb{407}
\end{equation}

     Now we consider $V_2(r)$.  To satisfy $0\leq q \leq \Lambda_c$, the domain of integration is not 
$(-\infty <q_n <\infty, 0\leq q_T \leq m_{\chi})$ in Eq.(\ref{401}), but 
$(-\varepsilon \leq q_n \leq \varepsilon, 0\leq q_T \leq \sqrt{\Lambda_c^2-\varepsilon^2})$ with $0<\varepsilon \ll 1$, i.e., 
\[ V_2(r)=\frac{Q^2m^2}{(2\pi)^2}\int_0^{\sqrt{\Lambda_c^2-\varepsilon^2}} dq_Tq_T\int_{-\varepsilon}^{\varepsilon}dq_n 
\frac{1-e^{iq_nr}}{q_n^2(q_n^2+q_T^2+m^2)}.  \]
Since the integrand is singular at $q_n=0$, we choose the anticlockwise path $\Gamma_{\varepsilon}$ in Fig.~2, and take the limit 
$\varepsilon \to 0$.  Then we obtain 
\begin{equation}
   V_2(r)=\frac{Q^2m^2}{(2\pi)^2}\int_0^{\Lambda_c} dq_Tq_T\frac{\pi r}{q_T^2+m^2}= \sigma_c r, \quad 
   \sigma_c=\frac{Q^2m^2}{8\pi}\ln\left(\frac{\Lambda_c^2 +m^2}{m^2}\right).  \lb{408}
\end{equation}
If the cut-off $\Lambda_c$ is replaced by $m_{\chi}$, $V_2(r)$ becomes the linear term in Eq.(\ref{102}).  

\begin{figure}
\begin{center}
\includegraphics{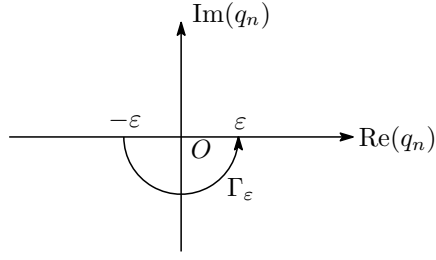}
\caption{The integration path $\Gamma_{\varepsilon}$ for $q_n$.}
\label{fig2}
\end{center}
\end{figure}

    Finally, neglecting additive constants, we find $V_3(r)$ becomes 
\begin{equation}
     V_3(r)=\frac{Q^2m^2}{2\pi^2}\int_0^{\Lambda_c} dq \frac{\sin qr}{qr}\frac{1}{q^2+m^2}.  \lb{409}
\end{equation}

     Thus the confining potential we propose is 
\begin{equation}
    V_c(r)=\sum_{k=1}^3 V_k(r)= -\frac{K_c}{r} +\sigma_c r 
     +\frac{Q^2m^2}{2\pi^2}\int_0^{\Lambda_c} dq \frac{\sin qr}{qr}\frac{1}{q^2+m^2}.  \lb{410}
\end{equation}
In addition to the Coulomb plus linear part, there is the term $V_3(r)$.

\section{Determination of parameters}

     Although we presented the potential $V_c(r)=\sum_{k=1}^3 V_k(r)$, the values of the parameters $Q^2$ and $m$ are unknown.  
To determine them, 
let us expand a potential $V(r)$ as 
\[ V(r)=V(r_c) + V'(r_c)(r-r_c) + \frac{V''(r_c)}{2}(r-r_c)^2 + \frac{V^{(3)}(r_c)}{3!}(r-r_c)^3 +\cdots.  \]
We assume there is a true confining potential $V_T(r)$.  Then we require the Cornell potential $V_{CL}(r)$ fits well to 
$V_T(r)$ at a point $r=r_c$.  This is achieved by choosing $(K, \sigma)$ and the constant $V(r_c)$ appropriately.  
\footnote{To determine the three parameters, it is natural to choose appropriate three points $r_k\ (k=1,2,3)$.  However, 
in the next step, we must determine $V_c$ so as to the difference between $V_c$ and $V_{CL}$ becomes minimum.  
Since $V_3$ in $V_c$ contains $r$ in the integrand, it is difficult to determine $r_k$.}
We impose the conditions 
\begin{equation}
    V_{T}^{(n)}(r_c)=V_{CL}^{(n)}(r_c),\quad (n=0,1,2),   \lb{501}
\end{equation}
and determine  $(K, \sigma)$ and $V(r_c)$.

     Next, we require that the potential $V_c(r)$ fits well to this $V_{CL}(r)$ at $r=r_c$, and 
use the conditions 
\begin{equation}
    V_{CL}^{(n)}(r_c)=V_c^{(n)}(r_c),\quad (n=0,1,2,3).   \lb{502}
\end{equation}
We note, to determine the parameters $Q^2$ and $m$, the two conditions with $n=1,2$ are necessary.  However, to determine $r_c$, 
the condition with $n=3$ is required.

     From Eqs.(\ref{101}) and (\ref{410}), we have 
\begin{align*}
 &-\frac{r^3}{2}V_{CL}''(r)=K,\quad  \frac{1}{2r}\left(r^2V_{CL}'(r)\right)'=\sigma, \\
 &-\frac{r^3}{2}V_{c}''(r)=K_c-\frac{r^3}{2}V_{3}''(r), \quad 
 \frac{1}{2r}\left(r^2V_{c}'(r)\right)'=\sigma_c +\frac{1}{2r}\left(r^2V_3'(r)\right)'.  
\end{align*}
Since Eq.(\ref{502}) with $n=2$ gives $r_c^3V_{CL}''(r_c)=r_c^3V_c''(r_c)$, 
this condition becomes 
\begin{equation}
    K=K_{ef}(r_c),\quad K_{ef}(r_c)=K_c-\frac{r_c^3}{2}V_3''(r_c).  \lb{503}
\end{equation}
In the same way, the condition $\displaystyle \left(r_c^2V_{CL}'(r_c)\right)'=\left(r_c^2V_C'(r_c)\right)'$ leads to 
\begin{equation}
    \sigma = \sigma_{ef}(r_c),\quad \sigma_{ef}(r_c)=\sigma_c + \frac{1}{2r_c}\left(r_c^2V_3'(r_c)\right)',  \lb{504}
\end{equation}
and $\displaystyle \left(r_c^3V_{CL}''(r_c)\right)'=\left(r_c^3V_c''(r_c)\right)'$ becomes 
\begin{equation}
   \left(r_c^3V_3''(r_c)\right)'=0.  \lb{505}
\end{equation}

     Of course, the true potential $V_T(r)$ is unknown.  So, instead of the first step proposed in Eq.(\ref{501}), we 
choose appropriate values of $K$ and $\sigma$.  Thus we determine $V_c(r)$ as follows.  
Choosing the values of the scale parameter $\Lambda_{\mathrm{QCD}}$ and the off-diagonal gluon mass $M$, 
Eq.(\ref{304}) determines the cut-off $\Lambda_c$.  Then, substituting the values of $\Lambda_c$, $K$ and $\sigma$ 
into Eqs.(\ref{503})-(\ref{505}), the quantities $Q$, $m$ and $r_c$ are determined numerically.  
As an example, we choose the values 
\begin{equation}
    \Lambda_{\mathrm{QCD}}=0.2\ \mathrm{GeV},\quad M=1.2\ \mathrm{GeV}, \quad K=0.3,\quad \sigma = 0.18\ \mathrm{GeV}^2.  \lb{506}
\end{equation}
We note that the off-diagonal gluon mass $M\simeq 1.2$ GeV in the region of $r \gtrsim 0.2$ fm was obtained by using SU(2) lattice QCD 
in the maximal Abelian gauge \cite{as}.  The values $K\lesssim 0.3$ and $\sigma \lesssim 0.2$ GeV$^2$ come from lattice 
simulations \cite{bali}.

     Now using the values of $\Lambda_{\mathrm{QCD}}$ and $M$ in Eq.(\ref{506}), Eq.(\ref{304}) gives 
$\Lambda_c\simeq 2.03$ GeV.  
Next, we substitute this $\Lambda_c$ and $(K,\ \sigma)$ in Eq.(\ref{506}) into Eqs.(\ref{503})-(\ref{505}), and solve these equations.  
The details are 
explained in Appendix~B.  The results are 
\[ r_c\simeq 1.145\ \mathrm{GeV}^{-1}=0.226\ \mathrm{fm},\quad a=m^2/\Lambda_c^2\simeq 0.263, \]
and these values lead to 
\begin{equation}
     m=1.04\ \mathrm{GeV},\quad K_c=\frac{Q^2}{4\pi}=0.285,\quad \sigma_c=\frac{Q^2m^2}{8\pi}\ln\left(\frac{\Lambda_c^2 +m^2}{m^2}\right)=0.242\ \mathrm{GeV}^2. \lb{507}
\end{equation}
Thus we obtain 
\begin{equation}
   V_c(r)=\sum_{k=1}^{3}V_k(r),\quad V_1(r)=-\frac{0.285}{r},\quad V_2(r)=0.242\cdot r,\quad V_3(r)=0.747\cdot S(r,0.263),  \lb{508}
\end{equation}
where 
\begin{equation}
   S(r,a)=\int_0^{\Lambda_cr}dx \frac{\sin x}{x}\frac{ar}{x^2+a(\Lambda_c r)^2}.  \lb{509}
\end{equation}
In Fig.3, the potentials $V_k(r), (k=1,2,3)$ and $V_c(r)$ are plotted.  Since $V_1(r)+V_3(r)$ is a substitute for the Yukawa potential, 
$V_1(r)+V_3(r)\approx 0$ for $r\gtrsim 0.35\ \mathrm{fm}$ is reasonable.  
Using the values of $(K,\ \sigma)$ in Eq.(\ref{506}), $V_{CL}(r)$ becomes 
\begin{equation}
    V_{CL}(r)=-\frac{0.3}{r} + 0.18\cdot r.   \lb{510}
\end{equation}
Eqs.(\ref{508}) and (\ref{510}) are plotted in Fig.~4.  Since $V_c(r)$ is fitted to $V_{CL}(r)$ at 
$r_c\simeq 0.226\ \mathrm{fm}$, they fit very well for $0.1\ \mathrm{fm}\lesssim r \lesssim 0.4\ \mathrm{fm}$.  However, when $r$ becomes large, 
as $V_1(r)+V_3(r)\approx 0$ and $\sigma_c>\sigma$, $V_c(r)>V_{CL}(r)$ holds for $r \gtrsim 0.4\ \mathrm{fm}$.  
In the same way, $K_c<K$ leads to $V_c(r)>V_{CL}(r)$ for $r< 0.09\ \mathrm{fm}$.

\begin{figure}
\begin{center}
\includegraphics[width=0.6\linewidth]{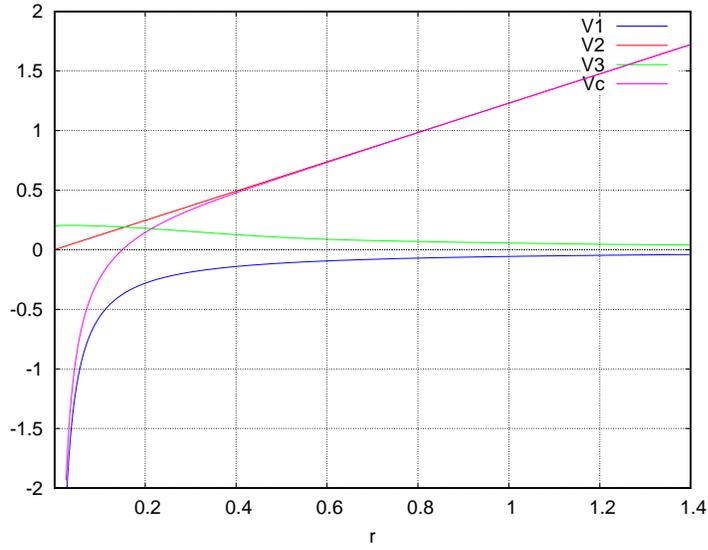}
\vspace{0.8cm}
\caption{The potentials $V_1, V_2, V_3$ and $V_c=\sum_{k=1}^3V_k$ in Eq.(\ref{508}).  The unit of $r$ is fm, and the unit of the potentials is GeV.}
\label{fig3}
\end{center}
\end{figure}

\begin{figure}
\begin{center}
\includegraphics[width=0.6\linewidth]{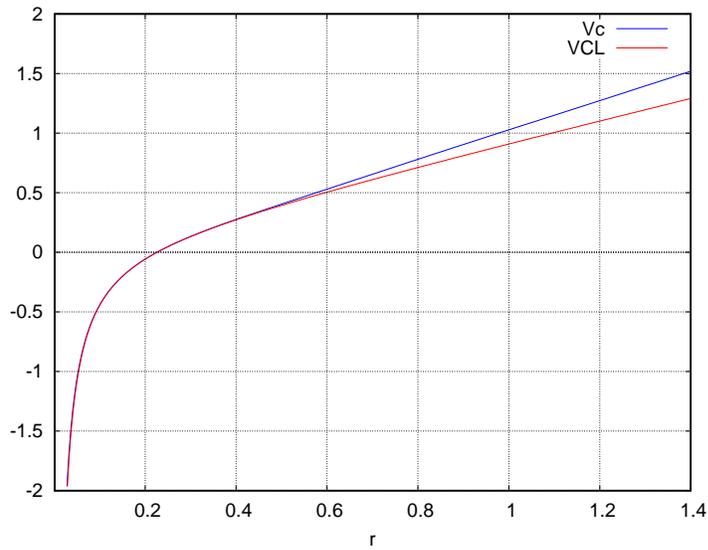}
\vspace{0.8cm}
\caption{The potentials $V_c$ in Eq.(\ref{508}) and $V_{CL}$ with $(K=0.3, \sigma=0.18\ \mathrm{GeV}^2)$.  Additive Constants are chosen 
to become $V_c(r_c)=V_{CL}(r_c)=0$, and the unit of $r$ is fm.}
\label{fig4}
\end{center}
\end{figure}

\section{The scales $R_0$ and $\tilde{R}_0$}

     In Sect.~5, the scale $r_c\simeq 0.226\ \mathrm{fm}$ appears.  On the other hand, considering the force $-V'(r)$, 
the intermediate scale $R_0$, which satisfies 
\begin{equation}
     r^2 V'(r)|_{r=R_0} = 1.65, \quad R_0\simeq 0.5\ \mathrm{fm},  \lb{601}
\end{equation}
was proposed \cite{so}.  In successful potential models, this relation holds fairly well.  For example, 
the $V_{CL}(r)$ with $(K=0.52,\ \sigma=0.183\ \mathrm{GeV}^2)$ \cite{cor2} gives $R_0=0.49\ \mathrm{fm}$.  
If we substitute $V_c(r)$ in Eq.(\ref{508}) into Eq.(\ref{601}), we obtain 
\begin{equation}
 K_{ef}(R_0)+R_0^2\sigma_{ef}(R_0)=0.285+0.242 R_0^2  +0.747R_0^2\cdot H(R_0,0.263)=1.65, \lb{602}
\end{equation}
where 
\begin{equation}
     H(r,a)= \int_0^{\Lambda_c r} dx \left(\cos x- \frac{\sin x}{x}\right)\frac{a}{x^2+a(\Lambda_c r)^2}  \lb{603}
\end{equation}
comes from $V_3'(r)$.  Eq.(\ref{602}) gives the solution $R_0=0.51\ \mathrm{fm}$.  
We note, $V_{CL}(r)$ with $(K=0.3,\ \sigma=0.18\ \mathrm{GeV}^2)$ in Sect.~5 gives the larger value $R_0=0.54\ \mathrm{fm}$.

\begin{figure}
\begin{center}
\includegraphics[width=0.6\linewidth]{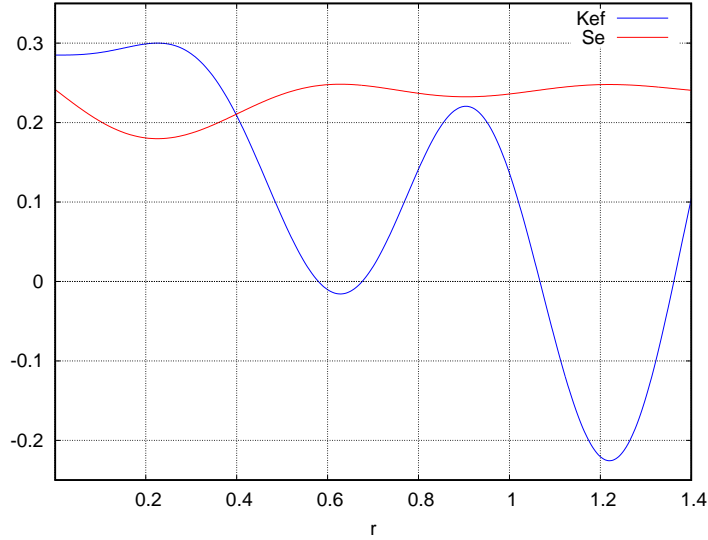}
\vspace{0.8cm}
\caption{The effective Coulomb coupling $K_{ef}(r)=K_c-\frac{r^3}{2}V_3''(r)$ 
and the effective string tension $S_e(r)=\sigma_{ef}(r)=\sigma_c + \frac{1}{2r}\left(r^2V_3'(r)\right)'$.  
The unit of $r$ is fm.}
\label{fig5}
\end{center}
\end{figure}

     To see the meanings of these scales, the effective Coulomb coupling $K_{ef}(r)=K_c-\frac{r^3}{2}V_3''(r)$ in Eq.(\ref{503}) 
and the effective string tension $\sigma_{ef}(r)=\sigma_c + \frac{1}{2r}\left(r^2V_3'(r)\right)'$ in Eq.(\ref{504}) are plotted in Fig.~5.  
We find $K_{ef}(r_c)=0.3$ is the maximal value and $\sigma_{ef}(r_c)=0.18\ \mathrm{GeV}^2$ is the minimal value.  
Namely $r_c$ is the position where $K_{ef}(r)$ 
is maximum and $\sigma_{ef}(r)$ is minimum.  We also notice that $K_{ef}(r)\approx 0$ at $r\approx 0.58\ \mathrm{fm}$, and 
$0.23\ \mathrm{GeV}^2<\sigma_{ef}(r)<0.25\ \mathrm{GeV}^2$ for $r\gtrsim 0.5\ \mathrm{fm}$.  

     In Fig.~6, $r^2V_c'(r)=K_{ef}(r)+r^2\sigma_{ef}(r)$ and $r^2\sigma_{ef}(r)$ are plotted.  
We find that $r^2V_c'(r)$ satisfies $r^2V_c'(r)\approx K_{ef}(r)$ for $r<0.2\ \mathrm{fm}$, and $r^2V_c'(r)\approx r^2\sigma_{ef}(r)$ 
for $r\gtrsim 0.5\ \mathrm{fm}$.  
Namely, the force-related quantity $r^2V_c'(r)$ 
is almost saturated with the string part $r^2\sigma_{ef}(r)$ above $R_0\simeq 0.5\ \mathrm{fm}$.  
Especially, as $K_{ef}(r)\approx 0$ at $r\approx 0.58\ \mathrm{fm}$, we find  
\[  r^2 V_c'(r)|_{r=\tilde{R}_0} \simeq r^2\sigma_{ef}(r)|_{r=\tilde{R}_0}= 2.13, \quad \tilde{R}_0\simeq 0.58\ \mathrm{fm}.  \]

\begin{figure}
\begin{center}
\includegraphics[width=0.6\linewidth]{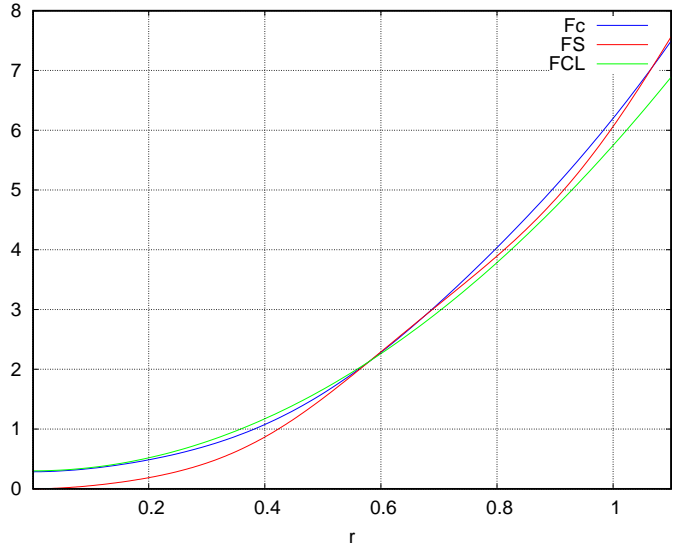}
\vspace{0.8cm}
\caption{$F_c(r)=r^2V_c'(r)=K_{ef}(r)+r^2\sigma_{ef}(r)$ and $FS(r)=r^2\sigma_{ef}(r)$.  For comparison, 
$FCL(r)=r^2V_{CL}'(r)=K+r^2\sigma$ with $(K=0.3,\sigma=0.212\ \mathrm{GeV}^2)$ is plotted.  The unit of $r$ is fm.}
\label{fig6}
\end{center}
\end{figure}

     Before closing this section, based on the above analysis, we present the potential $V_{CL}(r)$ that fits to $V_c(r)$ better in the region of $r>0.6\ \mathrm{fm}$.  
When $r$ is small, the Coulomb part $K_{ef}(r)$ dominates.  So keeping the condition Eq.(\ref{503}) intact, we set $K=0.3$.  
When $r$ becomes large, the string part dominates.  To determine the value of $\sigma$, it is reasonable to set the condition 
\begin{equation}
   r^2 V'(r)|_{r=\tilde{R}_0} = 2.13, \quad \tilde{R}_0\simeq 0.58\ \mathrm{fm}.  \lb{604}
\end{equation}
We find $V_{CL}$ with $(K=0.3, \sigma=2.12\ \mathrm{GeV}^2)$ satisfies Eq.(\ref{604}).  
We note this $V_{CL}$ satisfies Eq.(\ref{601}) as well.

     In Fig.~7, the potential $V_c(r)$ and 
\begin{equation}
     V_{CL}(r)=-\frac{0.3}{r} + 2.12\cdot r  \lb{605}
\end{equation}
are plotted .  
As we explained in Sect.~5, the behavior $V_c(r)>V_{CL}(r)$ comes about for large $r$.  
However Eq.(\ref{605}) fits fairly well for $r<1.2 \ \mathrm{fm}$.

\begin{figure}
\begin{center}
\includegraphics[width=0.6\linewidth]{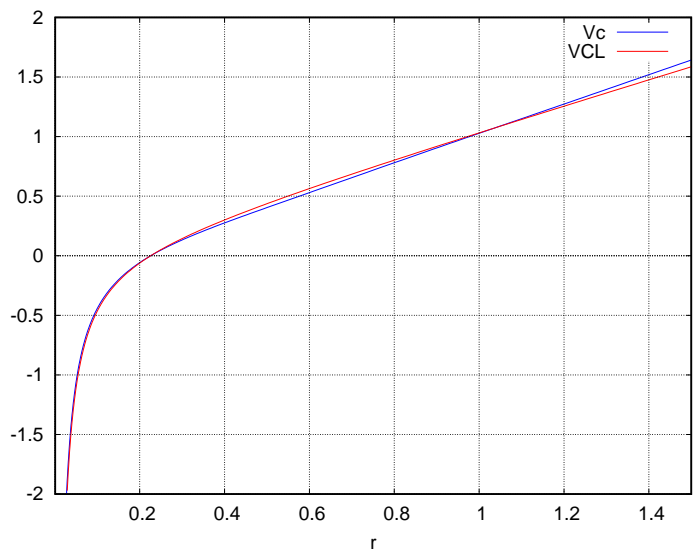}
\vspace{0.8cm}
\caption{The potentials $V_c$ in Eq.(\ref{508}) and $V_{CL}$ with $(K=0.3, \sigma=0.212\ \mathrm{GeV}^2)$.  The unit of $r$ is fm.}
\label{fig7}
\end{center}
\end{figure}

\section{Summary and comments}

     In this paper, we considered the SU(2) gauge theory, and studied the $Q\bar{Q}$ potential Eq.(\ref{215}).  
This potential is derived under the gauge field condensation.  In Refs.~\cite{suz, mts, sst, sst2}, 
the dual Ginzburg-Landau model, which describes the monopole condensation, leads to the potential.  

     In our approach \cite{hs,hs1}, the ghost condensation $v\neq 0$ appears, and it induces the VEV 
$G_{\mu\mu}^{(0)}(x,x)$, that is the lowest term of $\langle A_{\mu}^+A_{\mu}^- \rangle$.  
If we divide the diagonal gluon as $A_{\mu}^3=a_{\mu}^3+b_{\mu}$, the 
classical part $b_{\mu}$ acquires the mass 
$m=g\sqrt{2G_{\mu\mu}^{(0)}(x,x)} \simeq g\Lambda_{\mathrm{QCD}}/\sqrt{32\pi}$, 
whereas the quantum part $a_{\mu}^3$ is massless. 
The off-diagonal gluons $A_{\mu}^{\pm}$ acquire the mass $M$ through Eq.(\ref{303}).  
The low energy effective Lagrangian is Eq.(\ref{209}).  
As the classical solution $b_{\mu}$, 
we choose the electric monopole solution of the dual gauge field $\mcb_{\mu}$ \cite{hs1}.  
Then the propagator of $\mcb_{\mu}$ 
leads to the $Q\bar{Q}$ potential Eq.(\ref{215}).

    In calculating Eqs.(\ref{303}) and (\ref{403}), ultraviolet cut-off is necessary.  Above the scale 
$\Lambda_{\mathrm{QCD}}$, although $v$ vanishes, the masses $M$ and $m$ can exist.  
We assumed that these masses disappear above the cut-off $\Lambda_c$.  
Then the condition in Eq.(\ref{304}) and the confining potential $V_c(r)$ in Eq.(\ref{410}) are obtained.  
This potential has the linear potential $V_2(r)$ and, instead of the Yukawa potential, the Coulomb potential $V_1(r)$ and 
the additional term $V_3(r)$.

     Although we derived $V_c(r)$, there are unknown parameters.  To determine them, we chose the values 
presented in Eq.(\ref{506}).  Then, from Eq.(\ref{304}), the cut-off $\Lambda_c=2.03$ GeV was obtained.  
Next, assuming that the Cornell potential $V_{CL}(r)$ with $(K=0.3,\sigma=0.18\ \mathrm{GeV}^2)$ describes a true potential well 
at some point $r_c$, 
we required $V_c(r)\simeq V_{CL}(r)$ near $r=r_c$.  To realize this requirement, the conditions in Eq.(\ref{502}) are imposed.  
By solving these conditions, $r_c\simeq 0.226\ \mathrm{fm}$, the values of $m$ and $Q^2$ in Eq.(\ref{507}), and 
$V_c(r)$ in Eq.(\ref{508}) were obtained.

     There are two implicit scales $r_c$ and $R_0$ (or $\tilde{R}_0$) in $V_c(r)$.  To understand them, 
the effective Coulomb coupling $K_{ef}(r)$ in Eq.(\ref{503}) and the effective string tension 
$\sigma_{ef}(r)$ in Eq.(\ref{504}) were studied.  Since $V_3(r)$ contributes to them, they depend on $r$.  
At $r=r_c$, $K_{ef}(r)$ becomes maximum and $\sigma_{ef}(r)$ becomes minimum.  
For $r>R_0\simeq 0.5\ \mathrm{fm}$, $0.23\ \mathrm{GeV}^2<\sigma_{ef}(r)<0.25\ \mathrm{GeV}^2$ holds.

     If we consider the quantity $r^2V_c'(r)=K_{ef}(r)+r^2\sigma_{ef}(r)$, we find 
\[ r^2V_c'(r)\approx  \left\{
\begin{array}{ll}
 K_{ef}(r), &\quad (r<0.2\ \mathrm{fm}) \\
 r^2\sigma_{ef}(r), &\quad(r>0.5\ \mathrm{fm}).  
\end{array}
\right.
\]
Namely the main force between $Q$ and $\bar{Q}$ is the effective Coulomb force $-K_{ef}/r^2$ for $r<r_0$, and 
the effective string force $-\sigma_{ef}(r)$ for $r>R_0$.

     Although $V_c(r)$ was determined to fit to $V_{CL}(r)$ with $(K=0.3,\sigma=0.18\ \mathrm{GeV}^2)$ at $r=r_c$, 
it becomes larger than $V_{CL}(r)$ for $r>0.4\ \mathrm{fm}$.  
The Cornell potential is often used to fit lattice simulation data.  Can we find $V_{CL}(r)$ that fits to $V_c(r)$ better?  
To answer this question, we used the above scales.  
At $r_c$, Eq.(\ref{503}) was applied to determine $K$.  
To determine $\sigma$, we used Eq.(\ref{604}) at $\tilde{R}_0$.  
Then we obtained $V_{CL}(r)$ with $(K=0.3,\sigma=0.212\ \mathrm{GeV}^2)$.  This potential satisfies Eq.(\ref{601}) at 
$R_0$ as well, and fits fairly well in the region of $r< 1.2\ \mathrm{fm}$.

     We make three comments.

(1).\ In quark confinement, Abelian dominance \cite{ei} is expected.  
The lattice simulation in the maximal Abelian gauge shows that the linear part of the $Q\bar{Q}$ potential comes from 
the Abelian part \cite{ss}.  
In the present case, Abelian dominance is realized by the massive classical U(1) field $\mcb_{\mu}$.  
This field brings about the potential $V_c(r)$.  

(2)\ Although $V_c(r)$ comes from $\mcb_{\mu}$, to determine its behavior, information on the fields $a_{\mu}^A$ for 
$\Lambda_{\mathrm{QCD}}<\mu <\Lambda_c$ is necessary.  The low energy Lagrangian (\ref{209}) will be 
modified in this region.  However, as the first approximation, we used Eq. (\ref{209}) and assumed 
the cut-off $\Lambda_c$ for $m$.

(3).\ The values of $r_c$, $\tilde{R}_0$, $m$ and $Q^2$ in Eq.(\ref{507}) depend on the choice of Eq.(\ref{506}).  However the 
existence of these scales, and the behavior of $V_c(r)$, e.g., 
$\lim_{r\to \infty}V_3(r)=0$, $\sigma_c>\sigma$, $K_c<K$, and $V_c(r)>V_{CL}(r)$ for large $r$, are unchanged.

\appendix

\section{Propagator $G_{\mu\nu}^{(0)}(y,x)$ for $A_{\mu}^a$}

     Referring to Eqs.(\ref{202}) and (\ref{209}), we consider the Lagrangian with the 
massive gauge fields $A_{\mu}^{\pm}=(A_{\mu}^1\pm iA_{\mu}^2)/\sqrt{2}$:
\[ \sum_{a=1}^{2} \left\{\frac{1}{4}(\ptl\wedge A^a)_{\mu\nu}^2 +\frac{M^2}{2} (A_{\mu}^a)^2 
-\frac{\alpha_1}{2}(B^a)^2+B^a(\ptl_{\mu}A_{\mu}^a + \vp^a) + \frac{(\vp^a)^2}{2\alpha_2} \right\}.  \]
The fields $A_{\mu}$, $B$ and $\vp$ mix.  The inverse propagators of these 
fields are
\begin{equation}
  \bordermatrix{ &A_{\nu}& B & \vp \cr
  A_{\mu}& (p^2+M^2)P^T_{\mu\nu}+M^2P^L_{\mu\nu}& -ip_{\mu}& 0 \cr
  B& ip_{\nu}& -\alpha_1& 1 \cr
  \vp& 0& 1& \frac{1}{\alpha_2} }, \lb{a01}
\end{equation}
and the corresponding propagators are 
\begin{equation}
  \bordermatrix{ &A_{\nu}& B & \vp \cr
  A_{\mu}& \frac{1}{p^2+M^2}P^T_{\mu\nu}+  \frac{(\alpha_1+\alpha_2)}{\Xi}P^L_{\mu\nu}& 
  \frac{-ip_{\mu}}{\Xi} & 
  \frac{ip_{\mu}\alpha_2}{\Xi} \cr
  B& \frac{ip_{\nu}}{\Xi}& \frac{-M^2}{\Xi}& \frac{\alpha_2M^2}{\Xi} \cr
  \vp& \frac{-ip_{\nu}\alpha_2}{\Xi}& \frac{\alpha_2M^2}{\Xi}& \frac{\alpha_2(p^2+\alpha_1M^2)}{\Xi} },  \lb{a02}
\end{equation}
where $\Xi=p^2+(\alpha_1 +\alpha_2)M^2$, and 
\[ P^T_{\mu\nu}=\dl_{\mu\nu}-\frac{p_{\mu}p_{\nu}}{p^2}, \quad 
 P^L_{\mu\nu}=\frac{p_{\mu}p_{\nu}}{p^2}. \]

    Under the BRS transformation $\delta_B$, as $\delta_B \bar{c} =iB$ and $\delta_B B=0$, 
\[    \langle B B \rangle = -i\langle \delta_B(\bar{c} B)\rangle =0  \] 
holds, if the BRS symmetry is not broken spontaneously.  
When $M\neq 0$, to make 
\[ \langle B B \rangle = \frac{-M^2}{p^2+(\alpha_1 +\alpha_2)M^2}  \]
vanish, we choose $\alpha_2 \to \infty$.  
Then Eq.(\ref{a02}) becomes 
\begin{equation}
\bordermatrix{ &A_{\nu}& B & \varphi \cr
  A_{\mu}& \frac{1}{p^2+M^2}P^T_{\mu\nu}+  \frac{1}{M^2}P^L_{\mu\nu}& 
  0 & 
  \frac{ip_{\mu}}{M^2} \cr
  B& 0 & 0 & 1  \cr
  \varphi& \frac{-ip_{\nu}}{M^2}& 1 & \frac{p^2+\alpha_1M^2)}{M^2} }.   \lb{a03}. 
\end{equation}
Eq.(\ref{a03}) shows that $A_{\mu}$ mixes with $\vp$.  At the one-loop order, 
the ghost loop contributes to $\langle A_{\mu}A_{\nu}\rangle$, $\langle A_{\mu}\vp\rangle$ 
and $\langle \vp\vp\rangle$ as well.  However 
tachyonic behavior only appears in $\langle A_{\mu}A_{\nu}\rangle$.  
In Sect.~3, as the mixing like $\langle A_{\mu}\vp\rangle$ does not contribute, the propagator 
\[ G_{\mu\nu}^{(0)}(p)=\frac{1}{p^2+M^2}P^T_{\mu\nu}+  \frac{1}{M^2}P^L_{\mu\nu} \]
in Eq.(\ref{a03}) is used.

\section{Solution of Eqs.(\ref{503})-(\ref{505})}

     From Eq.(\ref{409}), we find the derivatives 
\begin{align*}
V_3'(r)&= Q^2\frac{m^2}{2\pi^2}\int_0^{\Lambda_c} dq \left(\frac{\cos qr}{r} - \frac{\sin qr}{qr^2}\right)\frac{1}{q^2+m^2},\\      
V_3''(r)&= Q^2\frac{m^2}{2\pi^2}\int_0^{\Lambda_c} dq \left(2\frac{\sin qr}{qr^3}-2\frac{\cos qr}{r^2} - \frac{q\sin qr}{r}\right)\frac{1}{q^2+m^2},\\ 
\left(r^3V_3''(r)\right)'&=Q^2\frac{m^2}{2\pi^2}\int_0^{\Lambda_c} dq \left(-q^2r^2\cos qr\right)\frac{1}{q^2+m^2}. 
\end{align*}     
Then, introducing the variables $x=qr$ and $a=m^2/\Lambda_c^2$, Eqs.(\ref{503})-(\ref{505}) become 
\begin{align}
 & K=\frac{Q^2}{4\pi}\left[1+\frac{\Lambda_c^2}{\pi}r_c^2\left\{2H(r_c,a)+L(r_c,a)\right\}\right], \lb{b01} \\
 & \sigma=\frac{Q^2}{4\pi}\left\{ \frac{\Lambda_c^2}{2}a\ln \left(1+\frac{1}{a}\right)-\frac{\Lambda_c^2}{\pi}L(r_c,a)\right\},  \lb{b02} \\
 & G(r_c,a)=\int_0^{\Lambda_cr_c} dx \frac{x^2\cos x}{x^2+a(\Lambda_cr_c)^2}=0,  \lb{b03} 
\end{align}
where $H(r,a)$ is defined in Eq.(\ref{603}), and 
\begin{equation*}
  L(r,a)= \int_0^{\Lambda_cr} dx x\sin x\frac{a}{x^2+a(\Lambda_cr)^2}.  
\end{equation*}
By eliminating $Q^2$, Eqs.(\ref{b01}) and (\ref{b02}) leads to 
\begin{equation}
   F(r_c,a)=\sigma +2\sigma \frac{\Lambda_c^2}{\pi}r_c^2H(r_c,a)+\frac{\Lambda_c^2}{\pi}\left(\sigma r_c^2+K\right)L(r_c,a)
   -K\frac{\Lambda_c^2}{2}a\ln \left(1+\frac{1}{a}\right)=0.  \lb{b04}
\end{equation}

\begin{figure}
\begin{center}
\includegraphics[width=0.6\linewidth]{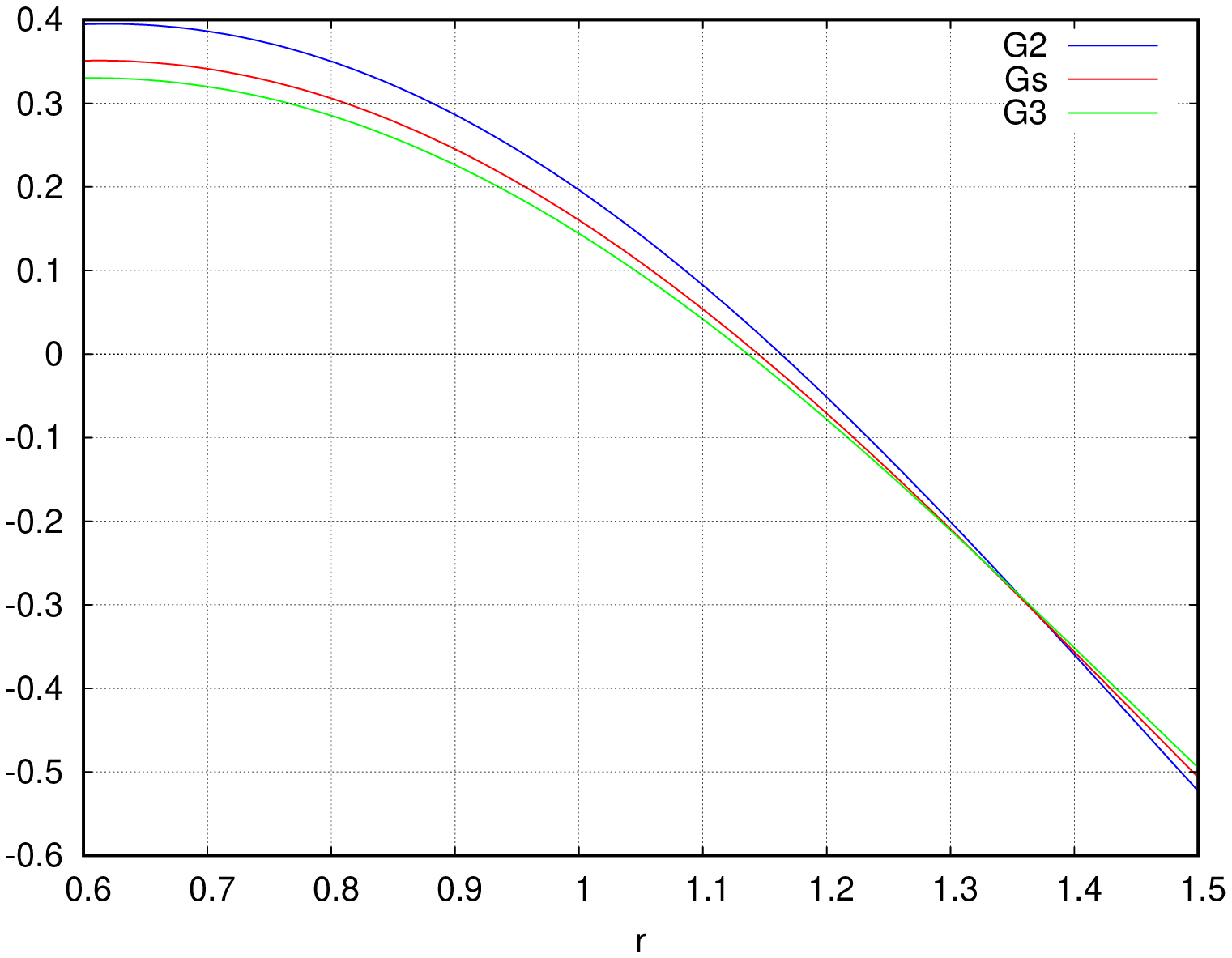}
\vspace{0.8cm}
\caption{The behavior of the function $G(r,a)$.  $G2=G(r,0.2)$, $G3=G(r,0.3)$ and $Gs=G(r,0.263)$ are plotted.  The unit of $r$ is GeV$^{-1}$.  
$Gs=0$ holds at $r\simeq 1.145\ \mathrm{GeV}^{-1}$.}
\label{fig8}
\end{center}
\end{figure}

\begin{figure}
\begin{center}
\includegraphics[width=0.6\linewidth]{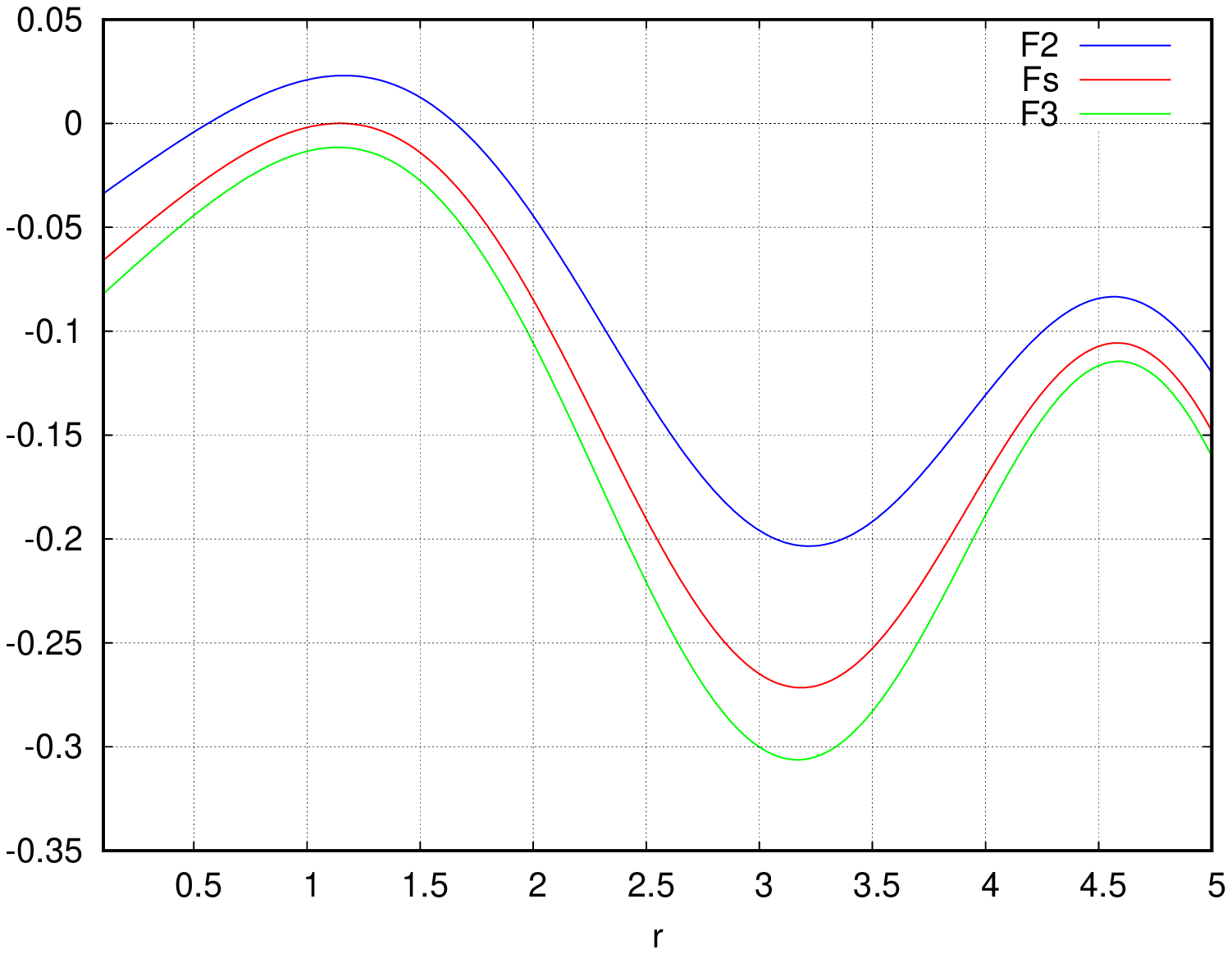}
\vspace{0.8cm}
\caption{The behavior of the function $F(r,a)$.  $F2=F(r,0.2)$, $F3=F(r,0.3)$ and $Fs=F(r,0.263)$ are plotted.  The unit of $r$ is GeV$^{-1}$.  
$Fs=0$ holds at $r\simeq 1.145\ \mathrm{GeV}^{-1}$.}
\label{fig9}
\end{center}
\end{figure}

     Now we substitute $K=0.3, \sigma=0.18\ \mathrm{GeV}^2$ and $\Lambda_c=2.03\ \mathrm{GeV}$.  
To solve Eq.(\ref{b03}) and Eq.(\ref{b04}) numerically, choosing $a=0.2, 0.263$ and $0.3$, 
$G(r,a)$ and $F(r,a)$ are plotted in Fig.~B1 and Fig.~B2, respectively.  
We find that $G(r_c,a)=0$ and $F(r_c,a)=0$ give the solutions $a\simeq 0.263$ and $r_c\simeq 1.145 \mathrm{GeV}^{-1}$.  

     Using these values, Eqs.(\ref{b01}) and (\ref{b02}) give 
\[ K_c=\frac{Q^2}{4\pi}\simeq 0.285, \quad \sigma_c=\frac{Q^2}{4\pi}\frac{\Lambda_c^2}{2}a\ln \left(1+\frac{1}{a}\right) \simeq 0.242\ \mathrm{GeV}^2.  \]
In the same way, using $S(r,a)$ in Eq.(\ref{509}), Eq.(\ref{409}) becomes 
\begin{equation}
   V_3(r)=2\frac{Q^2}{4\pi}\frac{\Lambda_c^2}{\pi}S(r,0.263)=0.747\cdot S(r,0.263). \lb{b05}
\end{equation}

\end{document}